\documentclass[reprint,aps,groupedaddress,superscriptaddress,twocolumn,assume,amsmath,floatfix,longbibliography]{revtex4-2}
\usepackage{natbib}

\usepackage{graphicx}
\usepackage[
  colorlinks=true,
  urlcolor=blue,
  linkcolor=blue,
  citecolor=blue
]{hyperref}
\usepackage{siunitx,braket,amssymb,gensymb,ulem,color}
\usepackage[utf8]{inputenc}

\usepackage{graphicx}
\usepackage{dcolumn}
\usepackage{bm}
\usepackage{ORCIDinREVTeX}

\begin{document}

\preprint{APS/123-QED}

\title{Recovering topological information of light by topological learning}

\author{Benquan Wang}
\thanks{These authors contributed equally to this work.}
\affiliation{Centre for Disruptive Photonic Technologies, School of Physical and Mathematical Sciences, Nanyang Technological University, Singapore 637371, Republic of Singapore}
\affiliation{Division of Mathematical Sciences, School of Physical and Mathematical Sciences, Nanyang Technological University, Singapore 637371, Republic of Singapore}
\author{Trishita Das}
\thanks{These authors contributed equally to this work.}
\affiliation{Centre for Disruptive Photonic Technologies, School of Physical and Mathematical Sciences, Nanyang Technological University, Singapore 637371, Republic of Singapore}
\author{Yuhan Peng}
\thanks{These authors contributed equally to this work.}
\affiliation{Division of Mathematical Sciences, School of Physical and Mathematical Sciences, Nanyang Technological University, Singapore 637371, Republic of Singapore}
\author{Tatjana Kleine}
\thanks{These authors contributed equally to this work.}
\affiliation{Centre for Disruptive Photonic Technologies, School of Physical and Mathematical Sciences, Nanyang Technological University, Singapore 637371, Republic of Singapore}
\affiliation{School of Physics, University of the Witwatersrand, Johannesburg 2050, South Africa}
\author{Shanshan Chang}
\affiliation{School of Electronic Science and Engineering, Xiamen University, Xiamen 361005, China}
\author{Jinhui Chen}
\affiliation{School of Electronic Science and Engineering, Xiamen University, Xiamen 361005, China}
\author{Mata‐Cervera Nilo}
\affiliation{Centre for Disruptive Photonic Technologies, School of Physical and Mathematical Sciences, Nanyang Technological University, Singapore 637371, Republic of Singapore}
\author{Chunyu Li}
\affiliation{Centre for Disruptive Photonic Technologies, School of Physical and Mathematical Sciences, Nanyang Technological University, Singapore 637371, Republic of Singapore}
\author{Kelin Xia}
\email{xiakelin@ntu.edu.sg}
\affiliation{Division of Mathematical Sciences, School of Physical and Mathematical Sciences, Nanyang Technological University, Singapore 637371, Republic of Singapore}
\author{Andrew Forbes}
\email{andrew.forbes@wits.ac.za}
\affiliation{School of Physics, University of the Witwatersrand, Johannesburg 2050, South Africa}
\author{Yijie Shen}
\orcid{0000-0002-6700-9902}
\email{Corresponding author: yijie.shen@ntu.edu.sg}
\affiliation{Centre for Disruptive Photonic Technologies, School of Physical and Mathematical Sciences, Nanyang Technological University, Singapore 637371, Republic of Singapore}
\affiliation{School of Electrical and Electronic Engineering, Nanyang Technological University, Singapore 639798, Republic of Singapore}

\begin{abstract}
\noindent \textbf{The evolution of modern-day communication networks towards enhanced capacity and robustness is driving interest in topological light wave as a next-generation information carrier. However, harnessing its potential is hindered by the lack of an effective topological reader, particularly when strong disorder renders conventional readout intractable. Here we address this by introducing TOPO$^{2}$, an artificial intelligence framework that leverages on topological learning to recover topological information from light. We demonstrate it on highly distorted optical fields with embedded Skyrmion topologies, showing that we can extract the topological numbers in a single shot with just a conventional camera.  We showcase its capability in a TOPO$^{2}$-based communication scheme, sending and recovering information encoded in a topological alphabet, confirming that it outperforms conventional approaches. Our connection of topological photonics with topological AI as a powerful paradigm for robust high-capacity information encoding and processing opens a new path towards topological information optics.
} 
\end{abstract}
\maketitle
\section*{Introduction}\label{sec1}

\noindent Light has many parameters and degrees of freedom to shape~\cite{forbes2021structured,he2022towards,rubinsztein2017roadmap,forbes2025progress}, that of emergent topological structures, e.g., optical skyrmions or skyrmionic beams~\cite{shen2024optical,shen2025free,chen2025more,ornelas2024non,liu2026broadband,ma2025nanophotonic}, has emerged as a promising information carrier for next-generation communication networks due to their inherent stability against continuous perturbations upon propagation~\cite{wang2024topological,guo2026topological,ornelas2025topological,zhang2025topological,ma2026high,liu2022disorder}. However, the practical utility of these topological states is hindered by the challenge of decoding information in complex environments~\cite{chen2026programmable,wang2025perturbation,zhang2026skyrmions,wan2023ultra,yang2025optical}. Because the topological stability was only endorsed in small enough continuous perturbations, the degradation of light in strong disorder with disconnected speckles appears to destroy the topology.  In addition, the standard decoding protocol of skyrmionic beams relies on Stokes polarimetry~\cite{peters2026extracting,shen2022generation,jia2025electrically}, a multi-shot process that works in controlled environments but becomes impractical in dynamic or strongly disordered media where the amplitude, phase, and polarization information are strongly mixed and scrambled into unrecognizable speckle patterns. Recent work has used a diffractive optical neural network to infer the skyrmion number from the OAM indices of two orthogonal polarization components~\cite{bezuidenhout2026deep}. However, recovering source topological information directly remains challenging, especially when the corresponding polarization texture is no longer identifiable at the receiver.

These limitations have motivated the adoption of artificial intelligence (AI) as a data-driven tool for recovering source optical information in complex environments \cite{zhang2025structured,fan2024dispersion,zhu2023harnessing,wu2025intelligent}. Deep learning approaches have demonstrated the ability to recognize structured light fields and speckle patterns without explicit analytical reconstruction \cite{raskatla2022speckle,Badavath2025SinglePixel,badavath2025mapping,das2024astigmatic,Sharma2024Nanostructures}. However, most existing AI-based optical decoding protocols rely on local intensity features or low-order statistical correlations and do not explicitly extract the underlying physical invariants of the optical field \cite{zhang2025structured,wang2022deep,zhang2024spatial,giordani2020machine,wang2024retrieving}. As a result, their robustness and generalization degrade under strong disorder, particularly when the optical signal is severely scrambled \cite{liu2023picophotonic,chi2024robust}. Moreover, the internal representations of many high-capacity neural network models remain difficult to interpret scientifically, making it unclear whether their predictive performance arises from physically meaningful field correlations \cite{li2025kolmogorov} or from other latent statistical features of the data.

Here, we introduce TOPO$^{2}$, a topology-enhanced scientific AI framework with mathematical interpretability that retrieves topological information of light directly from single-shot intensity measurements. By explicitly aligning the learning task with the skyrmion number as a global topological invariant of the polarization texture, TOPO$^{2}$ enables the direct and reliable recovery of topological information from scattered speckle patterns, allowing robust identification of optical skyrmion states and information transmission through strongly disordered channels without requiring Stokes polarimetry or reconstruction of the full vector optical field. This work establishes the first integration of topological light with topology-enhanced learning, elevating topology from a design principle for structured optical fields to a unifying framework governing both the encoding and decoding of optical information.

\begin{figure*}[t]
    \centering
    \includegraphics[width=\textwidth]{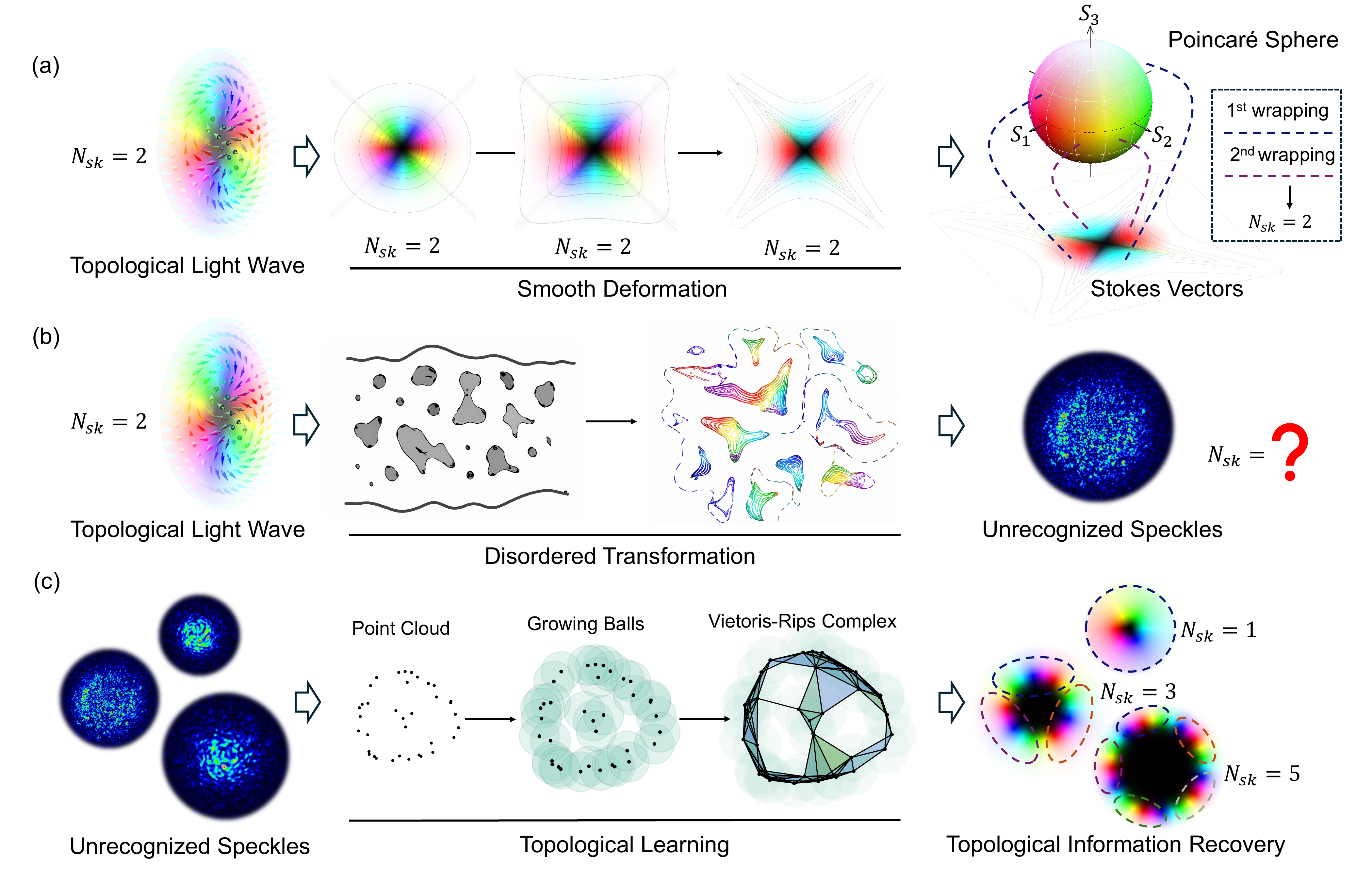}
    \caption{\textbf{Topology-enhanced AI for recovery of topological information.} (a) Topological light fields encode information in their skyrmion numbers ($N_{\mathrm{sk}} = 2$), but extracting the topology is expensive in both a measurement and computational sense.  (b) The problem becomes intractable after propagation through a strongly scattering disordered medium which scrambles the signal into visually indistinguishable speckle intensity patterns, preventing recovery of the underlying topology using conventional observables and machine learning approaches. (c) Our topology-enhanced AI learning framework (TOPO$^{2}$) maps these patterns into topology-aware representations, enabling direct recovery of the topological states from single-shot intensity measurements.}
    \label{fig:fig1}
\end{figure*}


\begin{figure*}[t]
    \centering
    \includegraphics[width=\textwidth, 
        clip]{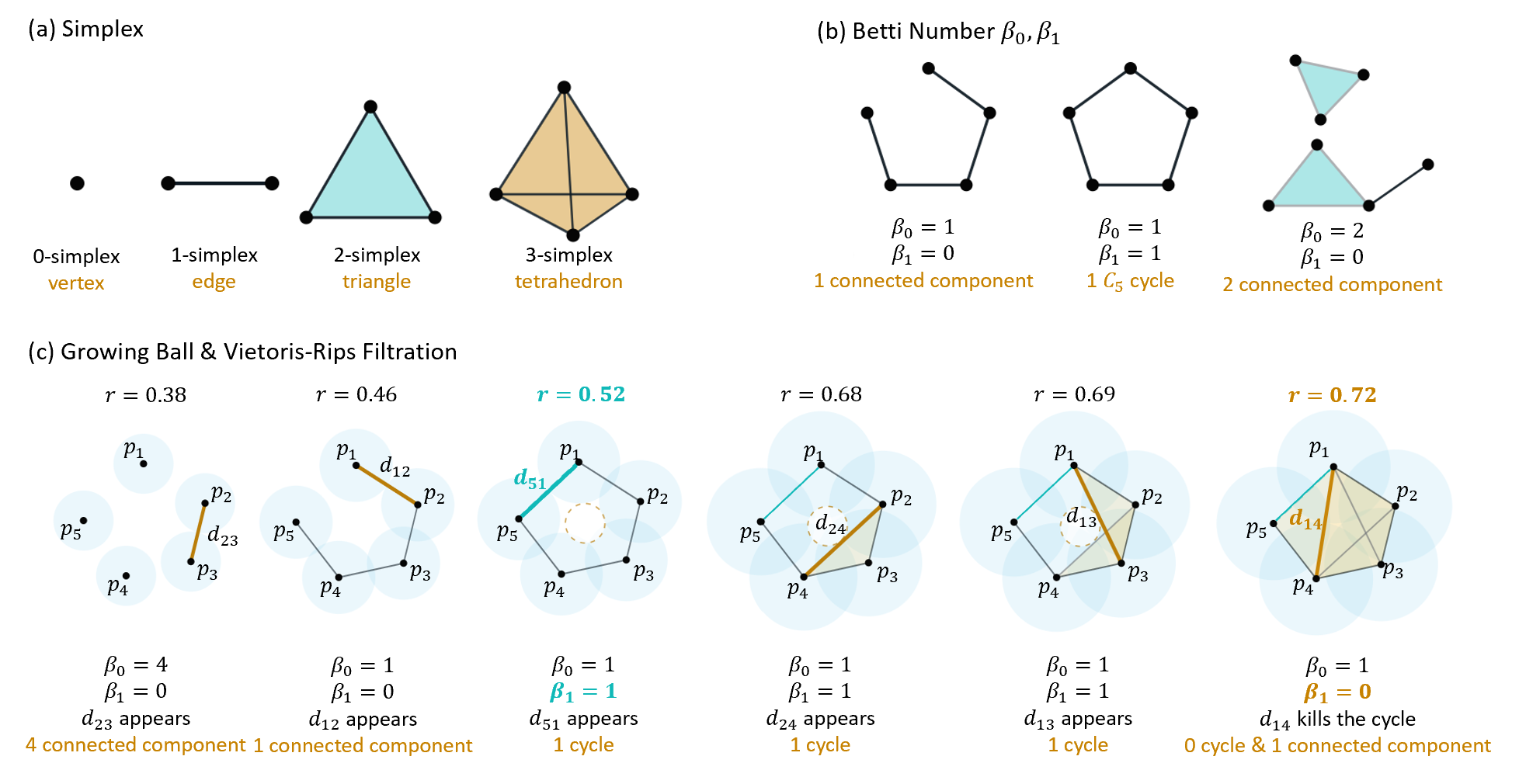}
    \caption{\textbf{Topological representation of information.} \textbf{(a)} A $k$ simplex is the convex hull of $k+1$ vertices: a vertex, edge, filled triangle, and tetrahedron for $k=0$--$3$. These are the units from which the geometry of the data is assembled. \textbf{(b)} Betti numbers count the structure features of that geometry: connected components ($\beta_0$) and independent loops ($\beta_1$)
    . Like the skyrmion number, they are unchanged by continuous deformation. \textbf{(c)} These features are extracted across all scales by growing a ball of radius $r$ each data point and linking points whose balls overlap. As $r$ increases, features appear and disappear: here a cycle is born at $r=0.52$, filled and died at $r=0.72$. How long each feature survives is the main descriptor TOPO\textsuperscript{2} uses to recover the topological state from a single-shot speckle pattern.}

    \label{fig:fig2}
\end{figure*}

\section*{Results}\label{sec2}

\subsection{Hidden topological information in disordered speckle}



Optical skyrmions, whether classical or quantum, carry a topologically invariant quantity, the skyrmion number $N_\mathrm{sk}$, which counts the number of times the normalized Stokes vector field $\mathbf{S}(\mathbf{r})$ wraps around the Poincar\'{e} sphere $S^2$ over the transverse plane. This number is conserved under continuous deformations of the field, making it a natural candidate for robust free-space information encoding. Skyrmionic textures are generated through the coherent superposition of two orthogonally polarized Laguerre--Gaussian spatial modes,
\begin{equation}
    |\Psi(\mathbf{r})\rangle = u^{l_1}_{p_1}(\mathbf{r})|L\rangle 
    + e^{i\theta_0} u^{l_2}_{p_2}(\mathbf{r})|R\rangle,
    \label{eq:superposition}
\end{equation}
where $u^l_p$ denotes a Laguerre--Gaussian mode with radial index $p$ and azimuthal index $l$. The mismatch in azimuthal phase winding when $|l_1| \neq |l_2|$ generates a spatially varying polarization, which can be assigned a skyrmion number,
\begin{equation}
    N_{sk} = \frac{1}{4\pi} \int \mathbf{S} \cdot 
    \left(
        \frac{\partial \mathbf{S}}{\partial x} 
        \times 
        \frac{\partial \mathbf{S}}{\partial y}
    \right) dx\, dy,
    \label{eq:skyrmion_number}
\end{equation}
which measures the signed solid angle subtended by the map $\mathbf{S}: \mathbb{R}^2 \rightarrow S^2$ on the Poincar\'{e} sphere. In this encoding scheme, information is written into $N_\mathrm{sk}$ at the source; upon propagation through complex or practical media, the 
polarization texture undergoes smooth deformation yet remains topologically intact, and $N_\mathrm{sk}$ is recovered at the receiver directly from the measured Stokes vector field via Eq.~(\ref{eq:skyrmion_number}) (Fig.~\ref{fig:fig1}a). 

However, whether the topological invariant of optical skyrmions remains recoverable after propagation through strongly disordered media has not been established, as no decoding platform exists for this regime, where conventional Stokes polarimetry becomes intractable once disorder scrambles the field into a fully developed polarization speckle pattern (Fig.~\ref{fig:fig1}b).

In this speckle regime, the ordered global organization of the Stokes mapping is destroyed.
The local Skyrme density,
\begin{equation}
\rho =
\mathbf{S}\cdot
\left(
\frac{\partial \mathbf{S}}{\partial x}
\times
\frac{\partial \mathbf{S}}{\partial y}
\right),
\label{eq:skyrme_density}
\end{equation}
remains finite but fluctuates in sign across the field. 
Statistical isotropy implies no preferred orientation on the Poincaré sphere, leading to a symmetric distribution of $\rho$ and a vanishing spatial mean over sufficiently large regions. 
Consequently, the net skyrmion number approaches zero \cite{Maxwell},

\begin{equation}
N_{sk} \rightarrow 0,
\label{eq:ns_zero}
\end{equation}
despite the persistence of local polarization twisting.



However, since the input and output fields are related by a deterministic transmission operator $\mathcal{T}$,
\begin{equation}
    \mathbf{E}_\mathrm{out} = \mathcal{T}\,\mathbf{E}_\mathrm{in},
    \label{eq:transmission}
\end{equation}
the measured intensity $I = |\mathcal{T} E_\mathrm{in}|^2$ is a nonlinear yet deterministic function of the incident field. The apparent randomness of the speckle stems from the complexity of the mode mixing, while the winding of the incident field imprints topology-dependent statistical correlations on the intensity distribution. Therefore, recovering this hidden structure requires a representation sensitive to the global geometric organisation of the speckle field rather than its local intensity statistics.

Here, we introduce TOPO$^{2}$, a topology-enhanced learning framework that recovers $N_\mathrm{sk}$ directly from single-shot speckle intensity patterns by extracting multi-scale topological features through persistent homology. These features capture the geometric organisation of the speckle field that is statistically inherited from the winding structure of the incident skyrmion state, and are insensitive to local intensity fluctuations carrying no topological information. Paired with downstream classifiers, TOPO$^{2}$ recovers $N_\mathrm{sk}$ from a single-shot intensity measurement without Stokes polarimetry (Fig.~\ref{fig:fig1}c).

\subsection{Aligning topological learning representation with topological light}

The premise of TOPO\textsuperscript{2} is that disorder scrambles the spatial organisation of the Stokes texture, so that the skyrmion number can no longer be obtained from it, while the winding of the incident field still conditions the statistical geometry of the transmitted intensity. Recovering this signature requires a representation whose primitives are topological, describing connectivity and loop structure rather than pixel-level intensity. TOPO\textsuperscript{2} builds such a representation using persistent homology~\cite{persistenthomology, persistenthomology2}, a method from topological data analysis that tracks the birth and death of topological features, such as connected components and loops, as the dataset is examined across multiple spatial scales.
It characterizes a dataset through global connectivity and cycle structure and is insensitive to local coordinate values, which makes it sensitive to the same kind of quantity the physics conserves. Topology therefore identifies the level of description at which the encoded information survives propagation.

TOPO\textsuperscript{2} operates on a discrete geometric object built from the optical speckle intensity field, summarized in Fig.~\ref{fig:fig2}. The building block is the $k$-simplex, the convex hull of $k+1$ vertices, which for $k = 0, 1, 2,3$ realizes a point, an edge, a filled triangle, and a tetrahedron (Fig.~\ref{fig:fig2}a). Simplices assemble into a simplicial complex whose topology is quantified by the Betti numbers
$\beta_{k}$, where $\beta_{0}$ counts connected components and $\beta_{1}$ count independent loops~\cite{books}
(Fig.~\ref{fig:fig2}b). 
Like $N_{\mathrm{sk}}$, Betti numbers are integer-valued topological invariants that characterize global structure and remain unchanged under continuous deformations.

A single complex fixes one length scale, whereas the geometric organization inherited from the skyrmion winding spans many scales. Persistent homology addresses this through a filtration. In the Vietoris–Rips construction, ball of radius $r$ is grown around each data point, and a $k$-simplex is added whenever the balls centered at its $k+1$ vertices intersect pairwise, equivalently when $d(v_i,v_j)\leq2r$ for every pair of vertices $v_i,v_j$, where $d(v_i,v_j)$ denotes the pairwise distance between $v_i$ and $v_j$~\cite{books} (Fig.~\ref{fig:fig2}c). Sweeping $r$ from small to large generates a nested sequence of complexes in which topological
features are born and later died. Fig.~\ref{fig:fig2}c traces one such sweep. As $r$ increases, five isolated points connect ($\beta_{0} = 5 \to 1$), a cycle is born when the pairwise distance $d_{51}$ is bridged ($\beta_{1} = 1$ at $r = 0.52$), the cycle persists across a range of scales, and it is filled when $d_{14}$ closes it ($\beta_{1} = 0$ at
$r = 0.72$). The lifetime of each feature, measured as the interval between its birth and death radius, provides the multi-scale descriptor that TOPO\textsuperscript{2} uses. Long-lived features report the geometric skeleton statistically imprinted by the incident topology, and short-lived features correspond to disorder-induced noise.

\begin{figure*}[t]
    \centering
    \includegraphics[width=\textwidth]{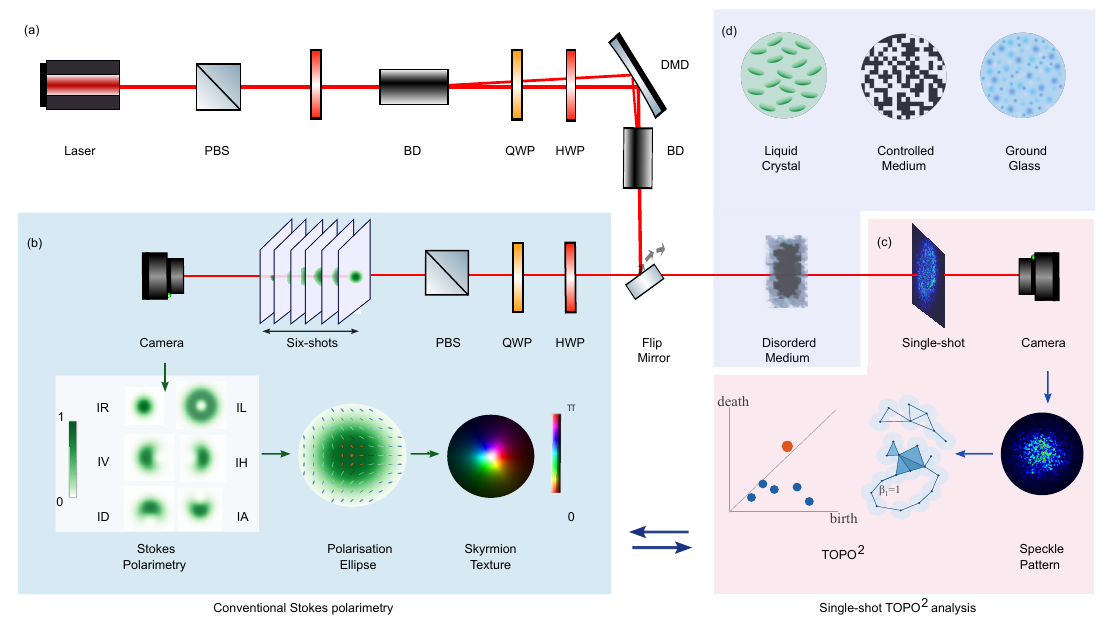}
    \caption{\textbf{Dual-mode optical platform for skyrmion generation and topological recovery.} (a) Optical skyrmions are generated by coherent superposition of two orthogonally polarized modes on a DMD. A flip mirror switches between (b) Stokes polarimetry for ground-truth characterization and (c) single-shot speckle recording through a disordered medium for TOPO$^{2}$ classification, which maps the optical speckle to a point cloud and extracts its persistent topology (here $\beta_1 = 1$) to recover the skyrmion number. (d) Disordered media consist of liquid crystals, digital amplitude distorting masks, and ground glass.}

    \label{fig:fig3}
\end{figure*}

\subsection{Single-shot topological information recovery platform without Stokes polarimetry}

To realize and validate this single-shot recovery paradigm experimentally, we construct an optical platform that supports two distinct operating modes within a shared beam path, as shown in Fig.~\ref{fig:fig3}. In the characterization mode, optical skyrmions with well-defined skyrmion numbers were synthesized and verified through full Stokes polarimetry, establishing ground-truth labels for each topological state. In the recovery mode, the same structured beams were directed through different disordered media, and the resulting speckle intensity patterns were recorded in a single camera exposure and fed directly into the TOPO$^{2}$ decoder, without reconstructing the vector optical field or resorting to Stokes polarimetry. Several disordered media were investigated, such as liquid crystals, ground glass and a variety of digital distortions designed to affect the phase or amplitude of the incoming light separately.

Skyrmion beams were generated by the coherent superposition of two orthogonally polarized Laguerre-Gaussian modes following Eq.~(\ref{eq:superposition}). A continuous-wave laser at wavelength 532~nm was spatially filtered, expanded, and collimated before illuminating a binary-amplitude DMD. The DMD surface was partitioned into two independently addressable regions, each encoding a binary hologram that defines one scalar spatial mode. After diffraction, the two beams were separated by a beam displacer (BD) and recombined through a $4f$ relay system, associating each mode with a well-defined circular polarization component $|L\rangle$ or $|R\rangle$. The relative azimuthal phase $\theta_0$ between the two channels was controlled to synthesize skyrmion states with target numbers $N_\mathrm{sk}$. Ground-truth characterization was performed by routing the beam into a Stokes polarimetry module comprising a sequence of wave plates, a polarizing beam splitter, and a CMOS camera, from which the full transverse Stokes vector field $\mathbf{S}(\mathbf{r})$ was reconstructed using traditional approaches and $N_\mathrm{sk}$ evaluated via Eq.~(\ref{eq:skyrmion_number}).

For the recovery experiment, a flip mirror redirected the skyrmion beam away from the polarimetry module and through the scattering stage. To demonstrate that single-shot topological recovery is not tied to any particular scattering mechanism, we employed three complementary classes of disordered media. First, a liquid-crystal (LC) layer functions as a thin, polarization-sensitive random scatterer whose spatially varying birefringence and refractive-index fluctuations induce strong mode mixing and depolarization, scrambling the incident structured field into a fully developed polarization speckle pattern (see Supplementary Note~1). Second, to establish the generality of the paradigm under a canonical, polarization-insensitive scatterer, we replaced the LC layer with a standard ground-glass diffuser, which produced well-developed scalar speckle through surface-roughness-induced random phase modulation. Third, we introduced a programmable perturbation medium whose degree of scrambling could be finely and reproducibly tuned, spanning weak to strong phase distortion together with partial and complete amplitude absorption. This controlled continuum of disorder isolated the dependence of topological recovery on distortion strength (see Supplementary Note~2). The transmitted intensity was recorded 30~cm downstream of the scattering stage using a CMOS camera (Thorlabs CS235CU, 1200~$\times$~1920 pixels), and the resulting single-shot intensity pattern was passed directly to TOPO$^{2}$ for topological classification. The two operating modes were otherwise identical in beam generation, ensuring that any distinction in the speckle patterns could be attributable solely to the topological state of the input.

\subsection{Label-free interpretable topological learning reveals hidden topology}

\begin{figure*}[t]
    \centering
    \includegraphics[width=\textwidth]{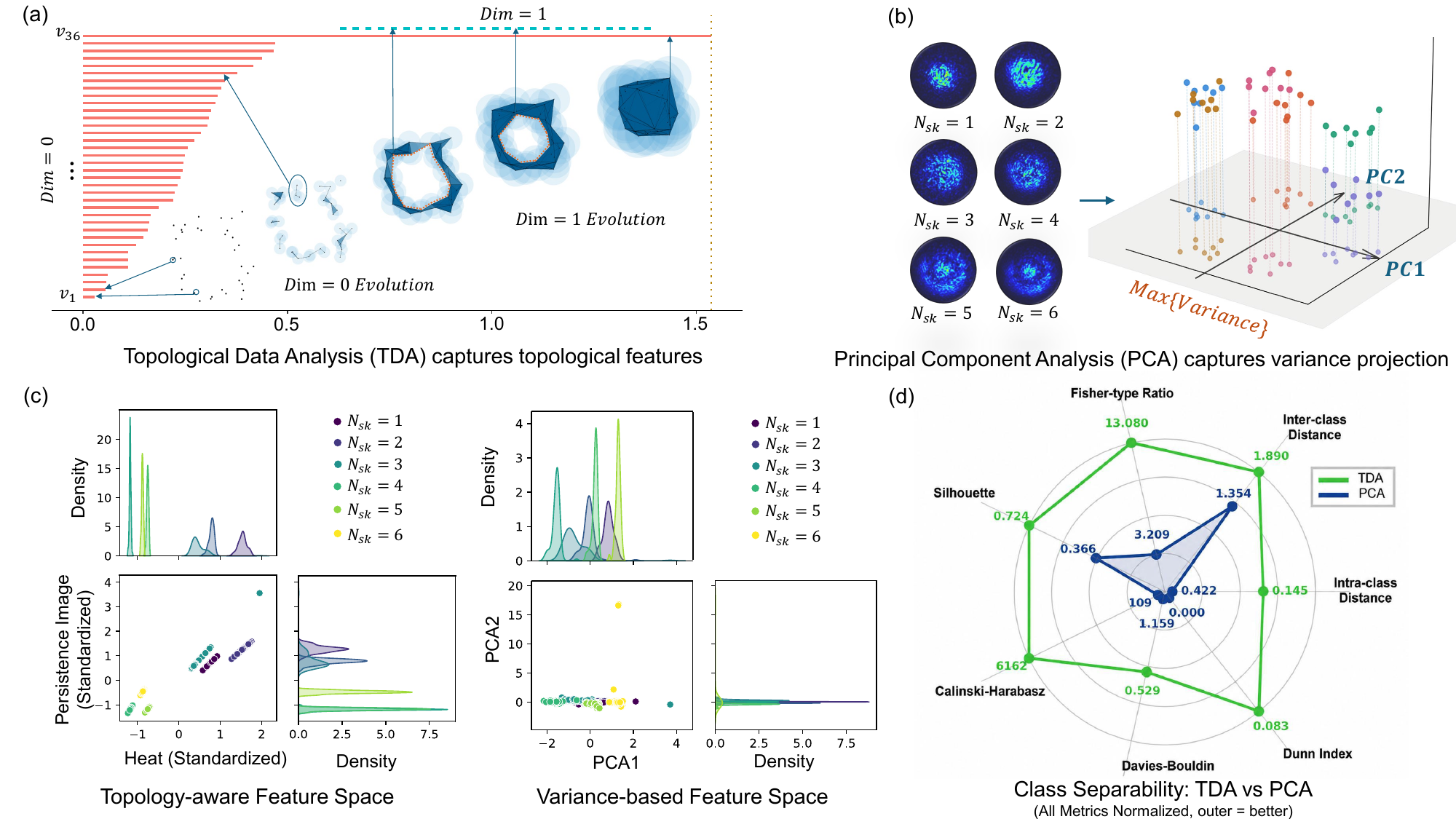}
    \caption{\textbf{Topological invariants remain separable under topology-aware representations. }    (a) Topological data analysis (TDA) extracts multi-scale geometric features from speckle intensity patterns via persistence diagrams, capturing connected components (dim 0) and loop structures (dim 1) across filtration scales. 
    (b)  Principal component analysis (PCA) projects speckle patterns onto principal components capturing dominant variance.
    (c) Comparison of the resulting feature spaces shows that the TDA representation exhibits distinct clustering by skyrmion number, whereas the PCA representation shows no clear classwise separation.
    (d) Quantitative comparison across multiple clustering metrics demonstrates consistent superiority of topology-aware representations over variance-based approaches.}
    \label{fig:fig4}
\end{figure*}

For each raw speckle pattern, we first convert it to a grayscale representation and resize it to $256\times256$ pixels, yielding a scalar intensity field $I(x,y)\in[0,255]$ on a uniform pixel grid. Each pixel is then represented as a point, with the pairwise distance between spatially neighboring points determined by their grayscale contrast, such that pixels with similar intensities lie close while sharp local variations remain separated in the resulting point cloud. Persistent homology is computed by applying the Vietoris--Rips filtration introduced above to this image-derived point cloud. Sweeping the filtration scale generates a nested sequence of simplicial complexes in which topological features are created (born) and subsequently merged or filled (die). In the context of speckle fields, dimension-$0$ ($k=0$) features correspond to connected components, while dimension-$1$ ($k=1$) features capture loop-like structures enclosed by intensity ridges, as shown in Fig.~\ref{fig:fig4}a. We track and summarize these multi-dimensional features by a persistence diagram~\cite{persistenthomology2, books},

\begin{equation}
\mathrm{PD}_k = \{(b_i^{(k)}, d_i^{(k)})\}_{i=1}^{m_k},
\end{equation}
where $b_i^{(k)}$ (birth) and $d_i^{(k)}$ (death) are the threshold values at which the $i$-th feature in dimension $k$ appears and disappears, satisfying $b_i^{(k)} < d_i^{(k)}$. Each point encodes the lifetime (persistence) $\ell_i^{(k)} := d_i^{(k)}-b_i^{(k)}$ of a topological feature; features with large persistence generally reflect robust geometric structures, whereas short-lived features are often associated with small-scale fluctuations or noise~\cite{stability}.

The persistence diagrams provide a multi-scale description of the geometry of the speckle intensity profile. To convert these variable-size diagrams into fixed-length numerical descriptors suitable for machine learning, we extract a comprehensive set of topology-aware features for each homological dimension $k\in\{0,1\}$~\cite{JMLR,mnist}.

\paragraph{Amplitude descriptors.}
Given a metric $\mathcal{M}$ on the space of persistence diagrams, the {amplitude is defined as the distance from $\mathrm{PD}_k$ to the empty diagram (diagonal multiset $\Delta$):
\begin{equation}
\mathrm{Amp}_{\mathcal{M}}(\mathrm{PD}_k) = d_{\mathcal{M}}(\mathrm{PD}_k, \Delta).
\end{equation}
We compute amplitudes under the following metrics:
\begin{itemize}
\item \textbf{Bottleneck}: $d_\infty(\mathrm{PD}_k, \Delta) = \max_i \frac{\ell_i^{(k)}}{2}$, capturing the single most persistent feature.
\item \textbf{Wasserstein} ($p$-th order): $d_{W,p}(\mathrm{PD}_k, \Delta) = \bigl(\sum_i (\ell_i^{(k)}/2)^p\bigr)^{1/p}$, aggregating all feature lifetimes.
\item \textbf{Landscape}: for each persistence pair, define the tent function
$\Lambda_i^{(k)}(t)
= \max\{0,\ell_i^{(k)}/2
- |t-(b_i^{(k)}+d_i^{(k)})/2|\}$.
The $j$-th persistence landscape is
$\lambda_j^{(k)}(t)
= j\text{-th largest}\{\Lambda_i^{(k)}(t)\}_i$.
The landscape amplitude is obtained from its $L^p$ norm.
\item \textbf{Heat kernel}: each persistence point is represented by a
Gaussian of standard deviation $\sigma$, together with a negative Gaussian
centered at its reflection across the diagonal. The resulting function is
$
\rho_{\sigma}^{(k)}(x)
=
\sum_i
\left[
G_\sigma\!\left(x-(b_i^{(k)},d_i^{(k)})\right)
-
G_\sigma\!\left(x-(d_i^{(k)},b_i^{(k)})\right)
\right],
$
and the heat-kernel amplitude is
$\mathrm{Amp}_{\mathrm{heat}}(\mathrm{PD}_k)
=\|\rho_{\sigma}^{(k)}\|_{L^p}$.
\item \textbf{Persistence image}: the persistence diagram is first
mapped from birth--death coordinates $(b_i^{(k)},d_i^{(k)})$ to
birth--persistence coordinates $(b_i^{(k)},\ell_i^{(k)})$. Gaussian
kernels are then centered at these points and sampled on a fixed grid to form a persistence image,
$
\pi_{\sigma}^{(k)}(x)
=
\sum_i w(\ell_i^{(k)})
G_\sigma\!\left(
x-(b_i^{(k)},\ell_i^{(k)})
\right).
$
The persistence-image amplitude is
$\mathrm{Amp}_{\mathrm{pimg}}(\mathrm{PD}_k)
=\|\pi_{\sigma}^{(k)}\|_{L^p}$.
\item \textbf{Betti curve}: $\beta_k(t) = |\{i : b_i^{(k)} \le t < d_i^{(k)}\}|$ counts the number of features alive at threshold $t$. The Betti-curve amplitude is
$\mathrm{Amp}_{\mathrm{Betti}}(\mathrm{PD}_k)
=\|\beta_k\|_{L^p}$.
\item \textbf{Silhouette}: the power-weighted silhouette is defined as
$
\phi_k(t)
=
\frac{\sum_i w_i^{(k)}\Lambda_i^{(k)}(t)}
{\sum_i w_i^{(k)}},
w_i^{(k)}=(\ell_i^{(k)})^q
$.
The silhouette amplitude is
$\mathrm{Amp}_{\mathrm{sil}}(\mathrm{PD}_k)
=\|\phi_k\|_{L^p}$.
\end{itemize}

\paragraph{Scalar descriptors.}
In addition, we extract the following topological descriptors:
\begin{itemize}
\item \textbf{Persistence entropy}: $H_k = -\sum_i p_i \log p_i$, where $p_i = \ell_i^{(k)} / \sum_j \ell_j^{(k)}$, quantifying the distribution of feature lifetimes as a Shannon entropy.
\item \textbf{Number of points}: $m_k = |\mathrm{PD}_k|$, the total count of topological features in dimension~$k$.
\item \textbf{Complex polynomial}: each persistence pair is mapped to
a complex root $z_i^{(k)}=b_i^{(k)}+\mathrm{i}d_i^{(k)}$, defining
$
P_k(z)=\prod_i\left(z-z_i^{(k)}\right).
$
The complex coefficients of $P_k$, represented by their real and
imaginary parts, provide a fixed-length descriptor of the persistence
diagram.
\end{itemize}






For each speckle image, the above descriptors are computed independently for $k=0$ and $k=1$, and concatenated into a single topology-aware feature vector. Importantly, this stage involves no labels or supervised training. Instead, the topology-aware descriptors provide an unsupervised feature representation of the data. 

\begin{figure*}[t]
    \centering
    \includegraphics[width=\textwidth]{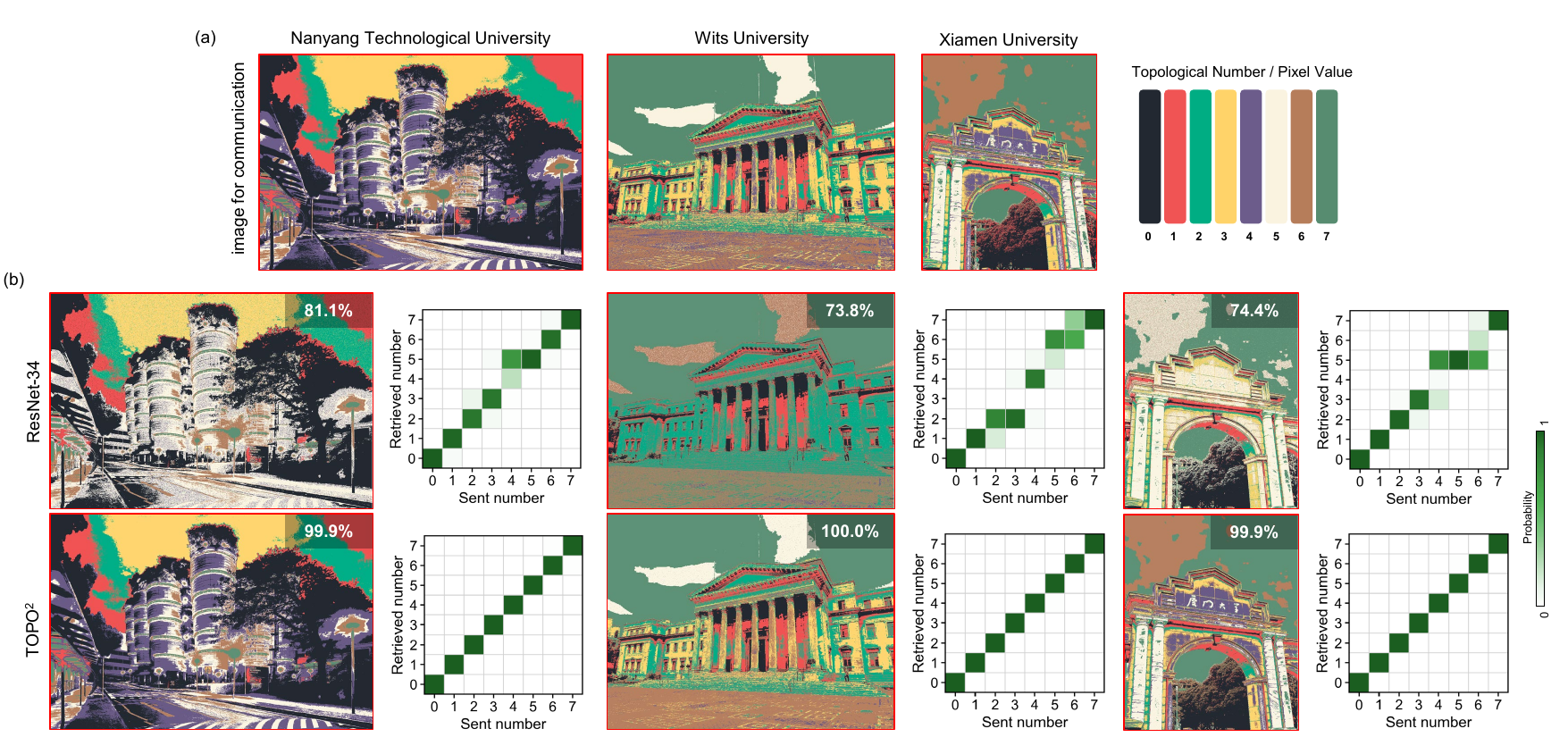}
    \caption{
    \textbf{Topology-based optical communication via encoding and decoding of topological states.} 
    (a) Topology encoding scheme. Image pixels are mapped to discrete skyrmion numbers, defining a topological alphabet. The resulting fields, upon propagation, generate speckle intensity patterns that retain topological signatures.
    (b) Topology-aware decoding from speckle. Encoded images are reconstructed from single-shot intensity using the TOPO$^2$ framework and compared with a ResNet-34 baseline. Confusion matrices reveal that topology-enhanced representations enable superior class discrimination and decoding fidelity across all tested cases. 
    }
    \label{fig:fig5}
\end{figure*}

To evaluate the structure of this representation, we first applied it to speckle patterns generated by a ground-glass diffuser as a general scattering medium, with skyrmion states of  
\begin{equation}
N_{sk} = 1\text{--}6.
\end{equation}
As a baseline, Principal Component Analysis (PCA) mathematically formalizes an orthogonal basis transformation that projects the high-dimensional data onto a lower-dimensional subspace spanned by the eigenvectors of the covariance matrix, ordered by their corresponding eigenvalues to sequentially maximize the retained variance. Applied directly to the raw speckle intensity patterns, PCA is expected to reveal any classwise structure that is linearly encoded in the intensity distribution, as shown in Fig.~\ref{fig:fig4}b. However, these two approaches yield fundamentally different outcomes: the variance-driven PCA projections exhibit severe statistical overlap, rendering the six skyrmion classes indistinguishable, whereas the topology-aware feature space organizes the same patterns into compact, well-separated clusters that are naturally ordered by skyrmion number, as shown in Fig.~\ref{fig:fig4}c. 



To quantify the class separability in the feature representation
, we evaluate the following standard metrics. Let $\mathbf{x}_j^{(c)}$ denote the $j$-th sample in class~$c$, $\boldsymbol{\mu}_c$ the centroid of class~$c$, and $n_c$ the class size. The intra-class distance measures the average within-class spread,
\begin{equation}
d_{\mathrm{intra}} = \frac{1}{C}\sum_{c=1}^{C} \frac{1}{n_c^2} \sum_{j,k} \|\mathbf{x}_j^{(c)} - \mathbf{x}_k^{(c)}\|,
\end{equation}
and the inter-class distance measures the average separation between class centroids,
\begin{equation}
d_{\mathrm{inter}} = \frac{2}{C(C-1)} \sum_{c < c'} \|\boldsymbol{\mu}_c - \boldsymbol{\mu}_{c'}\|.
\end{equation}
Their ratio defines the Fisher-type separation ratio $\mathcal{F} = d_{\mathrm{inter}} / d_{\mathrm{intra}}$, where higher values indicate stronger class separation relative to
within-class spread. In addition, we compute the Silhouette coefficient, the Calinski--Harabasz index, the Davies--Bouldin index, and the Dunn index, which together provide a comprehensive assessment of cluster compactness and separation~\cite{cluster}.

Across all seven metrics, the topology-aware representation consistently and substantially outperforms PCA and Fig.~\ref{fig:fig4}d summarizes the results. In particular, the Fisher-type ratio improves by a factor of $4.0\times$ (13.08 vs.\ 3.21), and the Calinski--Harabasz index improves by $56.5\times$ (6162 vs.\ 109), indicating that the topological features produce dramatically more compact intra-class clusters and wider inter-class separation than variance-based projections. Such topology-aware descriptors are obtained in an unsupervised manner and do not require labeled training data, which form a compact and interpretable representation of the speckle patterns and serve as feature vectors for downstream learning tasks.



\subsection{Encoding and decoding information with topology}

Having established that TOPO$^{2}$ yields an unsupervised feature space in which speckle patterns from different skyrmion states become naturally separable, we next asked whether this representation can support optical information transfer through disorder. To this end, we implemented a topology-based communication protocol in which information is encoded in the skyrmion number and decoded directly from single-shot speckle intensity after transmission through the ground-glass scattering channel. The receiver infers the transmitted topological state from the speckle alone, thereby converting topology from a property of the optical field into a directly usable information carrier.

The communication scheme is summarized in Figure~\ref{fig:fig5}a. Each image pixel is first quantized into a three-bit value and mapped onto a discrete skyrmion number ($N_{\mathrm{sk}} = 0$–$7$), defining an eight-symbol topological alphabet. This topological state determines the structured optical field launched into the disordered channel. After propagation through the disordered channels, the field is converted into a speckle pattern that appears visually random, yet still carries topology-dependent statistical structure. The decoding task is therefore not image reconstruction in the conventional sense but identification of the transmitted topological symbol from the scattered intensity pattern. Repeating this process pixel by pixel enables full image recovery.

Figure~\ref{fig:fig5}b presents the experimental demonstration on three representative images. In each case, the original scene is transformed into a topology-encoded image, transmitted through the disordered medium, and reconstructed from the received speckle using the trained TOPO$^{2}$ decoder, in which the topology-aware feature vectors are classified by a lightweight multilayer perceptron (MLP; see Supplementary Note~3.1). Across all three examples, the reconstructed images show high fidelity to the original encoded patterns. For the Nanyang Technological University landmark image, TOPO$^{2}$ achieves a decoding accuracy of 99.9\%, substantially outperforming a ResNet-34 baseline trained directly on raw speckle images, which achieves 81.1\%. Similar performance 
advantages are observed for the University of the Witwatersrand and Xiamen University landmarks (see Supplementary Note~3.2 for implementation details).

Beyond the ground-glass diffuser, we applied the same pipeline without modification to the liquid-crystal layer and to the programmable perturbation medium. The liquid-crystal layer is the more demanding case, since its spatially varying birefringence mixes the two polarization components and produces polarization speckle rather than the scalar speckle of the ground-glass diffuser. On this channel, TOPO$^{2}$ reaches an eight-class test accuracy of $93.25 \pm 1.13\%$, compared with $80.06 \pm 5.98\%$ for a ResNet-34 trained from random initialization. For the programmable medium, in which three skyrmion states ($N_{\mathrm{sk}} = 1, 3, 5$) were subjected to a range of controlled phase and amplitude distortions, the topological descriptors combined with a random forest yield an overall accuracy of 87.9\%, covering phase-only distortion of weak to strong strength, partial amplitude absorption, and complete amplitude absorption with up to 80\% of the mask opaque.

The performance gap originates from the TOPO$^{2}$ representation rather than from the choice of downstream classifier. Taking the liquid-crystal channel as an example, when the MLP is replaced by a support vector machine, XGBoost, or a random forest while the topological feature vectors are kept unchanged, the test accuracy remains within 87.6--93.3\%, consistently above the 80.1\% obtained by a ResNet-34 trained from random initialization on the raw speckle images (see Supplementary Note 3.3). The topological information is therefore already made explicit by the topological descriptors, and the classifier serves only to read it out. This is most evident in the comparison with a fine-tuned model based on ImageNet-pretrained ResNet-34 (see Supplementary Note 3.4), which benefits from both a much larger number of parameters and prior training on more than a million natural images, yet reaches $91.25 \pm 2.57\%$, still below the $93.25 \pm 1.13\%$ of TOPO$^{2}$. This indicates that the topological representation captures the physical structure carrying the encoded information so that a lightweight classifier suffices to read it out without relying on large-scale learned features.

\section*{Discussion}

Strong scattering through disordered media is conventionally assumed to obscure the 
physically meaningful structure of optical fields, scrambling the phase and polarization 
observables into randomized speckle patterns. Here, by introducing TOPO$^{2}$, a topology-enhanced artificial intelligence framework, we demonstrate that topological information encoded in structured light is not eliminated by disorder, where it is redistributed into the statistical geometry of the resulting speckle field and can be recovered directly from single-shot speckle intensity without reconstructing the optical field or resorting to Stokes polarimetry. This establishes a new paradigm for optical information transmission through complex media, in which information is written into topology and read out through a topology-aware representation. Furthermore, by operating on global geometric features rather than raw intensity distributions, TOPO$^{2}$ avoids the large parameter count and training cost of conventional convolutional networks. This computational lightness is not only practically efficient but also better aligned with physical implementation, making the framework naturally compatible with resource-constrained platforms of photonic computing. Together, these findings establish topology as a practical information resource in optics, laying a foundation for topological information optics.

\section*{Acknowledgements}
The authors acknowledge financial support from Singapore Ministry of Education (RG157/23, RT11/23); Singapore Agency for Science, Technology and Research (M24N7c0080, H25-MRO3489); Nanyang Technological University (Nanyang Assistant Professorship); Nanyang Technological University (Wallenberg--NTU Presidential Postdoctoral Fellowship); National Natural Science Foundation of China (12274357).

\section*{Author contributions}

B.W. conceived the idea and methodology. B.W. and Y.P. developed the algorithm. T.D., T.K., B.W., C.L., and N.M.-C. designed the experiment. T.D. and T.K. collected the experimental data. S.C. and J.C. fabricated the liquid-crystal sample.  B.W., T.D., and Y.P. wrote the original draft. Y.S., K.X., and A.F. supervised the research. All authors discussed the results and contributed to the manuscript.

\section*{Declaration of interests}
The authors declare no competing interests.

\section*{Data and code availability}
All data supporting the findings of this study are available within the paper and its supplementary information. The source data and code have been deposited in the Nanyang Technological University research repository under the accession code: \url{https://doi.org/10.21979/N9/4EJCDH}.

\bibliographystyle{unsrtnat}
\bibliography{sn-bibliography}

\end{document}